# Diffusion of Rb atoms in paraffin - coated resonant vapor cells

S.N. Atutov, F.A. Benimetskiy, A.I. Plekhanov, V.A. Sorokin, A.V. Yakovlev

1. *Institute of Automation and Electrometry of the Siberian Branch of the Russian Academy of Science, Koptyug Ave. 1, Novosibirsk, Russia, 630090*

2. *Novosibirsk State University, Pirogova street 2, Novosibirsk, Russia, 630090*

**Abstract.** We present the results of a study of the diffusion of Rb atoms in paraffin - coated resonant vapor cells. We have modeled the Rb diffusion both in the cell and in the coating, assuming that the main loss of Rb atoms is due to the physical absorption of the atoms by the glass substrate. It is demonstrated that the equilibrium atomic density in the cell is a monotonic function of the thickness of the paraffin coating: the density increases with an increase in the thickness of the coating. The diffusion coefficient for rubidium in paraffin thin films has been determined to be equal to $4.7 \cdot 10^{-7}$ cm$^2$/s. The results of the experiment might be useful for a better understanding of the details involved in the processes of the interaction of alkali atoms with a paraffin coating and atomic diffusion in resonant vapor cells.

## 1. Introduction

Many publications in the literature are devoted to experiments with optical resonant cells. This kind of cell is widely used in many experiments concerning the magneto-optical trapping of radioactive isotopes or rare atoms [1, 2], atomic clocks [3], magnetometers [4-6], fundamental symmetry studies [7], electromagnetically induced transparency [8], spin squeezing [9], long-lived entanglement [10] and quantum memory [11].

The major difficulty in any resonant cell experiments lies in the atomic vapor's interaction with the inner wall of the resonance cell. It is known that, as atoms collide with the surface, they are subject to an attractive potential whose range depends on the electronic and atomic structures of the surface and the atoms. A fraction of the atoms can be trapped in the attractive-potential well at the surface by the physical adsorption process. The adsorbed particles can be desorbed back into the gas phase from the surface after some interaction time, if the energy that has been imparted to them from the surface is enough to overcome the surface Van der Waal's attractive force. Atoms are lost to the wall through chemisorptions: they react chemically with the wall and are permanently removed from the vapor. In the trapping of radioactive atoms, the atoms are also lost to the wall through physical adsorption: the atom sticks to the wall for a long time and decays on the surface. In the case of trapping of rare atoms, a long interaction time leads to a much lower number of the atoms in the vacuum volume of an entire cell in respect to the number of absorbed atoms on the surface. This leads to a low efficiency in the collection of atoms in a magneto-optical trap. In experiments with polarized atoms, the long interaction time causes a high rate of the atomic spin relaxation on the glass walls of the cell.

The use of polymer non-stick (anti-relaxation) coatings has substantially overcome the difficulties noted in the above studies. This kind of polymer coating is characterized by a low surface potential that greatly reduces the atom – surface interaction time and the probability of the absorption of the atom by the surface in one collision. The anti-relaxation coating was suggested for the first time in [12] and studied in [13, 14]. Now, many publications deal with studies of different sorts of non-stick coatings with a view to overcoming the problem discussed above (see, [15] and references therein). Given the importance of the application of polymer coatings in optical experiments with the resonant excitation of atoms, extensive studies have been made of the polymers used (see, [16, 17] and references therein), and a large number of important and interesting results have been obtained.

Paraffin is widely used as an anti-relaxation coating in experiments involving alkali metal resonant cells, therefore a complex of the macroscopic processes, such as diffusion and optical pumping of the Rb atoms in a paraffin - coated resonant cell, is the main goal of the study. We present the results of an investigation into the variation in the density of Rb atoms as a function of the thickness of the paraffin coating, the results of the measurements of the diffusion coefficient of Rb atoms in a paraffin compound and the results of study of the optical pumping processes of the atoms in the paraffin-coated cell. These experimental studies are preceded by a discussion of a model of the rubidium diffusion and pumping in the paraffin - coated cell and by the definition of the relevant quantities.

## 2. Diffusion of Rb atoms in a paraffin-coated vapor cell

Obviously, a large sample of alkali atoms in a coated cell is of great importance to experiments with optical resonant cells. However, it has often been observed that in alkali metal resonant cells, the equilibrium of the atomic vapor density is 10 – 70 % lower than the vapor density in the vapor source of a cell that contains a piece of the metal used (see, [18] and references therein). It is generally believed that the observed suppression of the atomic vapor densities is attributable principally to a continuous loss of atoms due to the chemical reaction of the atoms with the polymer coating in the resonant cells. Indeed, the chemical reactions with silicon coatings and the resultant weak out-gassing have been reported [19 - 22]. It has also been determined that the chemical out-gassing produces a small amount hydrogen molecules (that are typical species of the background gas) and traces of silicon – containing species. It is obvious that the permanent loss of the atoms in cells causes a continuous flow of the atoms from the atomic source to a cell volume. If two each Rb atoms (involved in chemical interaction with the coating, rest gas or glass substrate) produce one hydrogen molecule, this continuous flow in cell volume $10 cm^3$ (typically $10^{11} s^{-1}$) would produce hydrogen gas with permanently increased pressure about 10 Torr per year. On the other hand, it is well known that collisions of hydrogen molecules with Cs atoms shift their hyperfine resonance frequencies by 1,9 kHz / Torr [23]. Hence, permanently elevated background gas pressure would lead to long-term drifts of hyperfine resonance frequencies. But this, presumably very large, frequency shift has not yet been observed, even in a 40-year-old paraffin – coated Rb cell: this cell shows comparable microwave frequency shifts to that of more recently manufactured cells [24]. From these facts, it can be concluded that both the pressure of the chemically produced hydrogen and, therefore, the rate of the chemical bonding of the Rb atoms are both negligible with respect to other losses in coated resonant cells.

We believe that the main loss of atoms in a properly cleaned and cured cell is due to the physical adsorption of the atoms on the glass substrate of the coating. This process does not increase the background gas pressure in the cell. Note that the number of atoms absorbed on the glass substrate can be very large. In fact, it was shown that over a period of about a week at the high temperature of 94 °C, a surface of Pyrex glass collects 6–7 equilibrium mono-layers of Rubidium [25]. It is clear that, at a room temperature when the equilibrium Rubidium vapor pressure is lower, it takes a much longer time to collect the same number of mono-layers. At a temperature of 94 °C, the equilibrium Rubidium vapor pressure is equal to $4 \times 10^{12}$ $cm^{-3}$; at a temperature 20 °C it is equal to $5 \times 10^9$ $cm^{-3}$ [26]. Therefore, at room temperature it would take more than 20 years to collect the same number of equilibrium mono-layers of Rubidium and obtain a truly steady-state atomic density in a Rb resonant cell.

Let us consider a buffer-gas-free resonant spherical cell consisting of a glass bulb (radius $R$) and cylindrical tube (internal radius $r$ and length $L$), connecting through a valve to an appendix with a piece of metallic Rb. The inner surface of the bulb, tube and valve, excluding appendix, have been covered with a paraffin coating (average thickness $l$). A gradient of the density of the atoms from the appendix, where the density ($n_0$) is maximum, to the glass substrate surface, where the density is zero ($n = 0$), causes a permanent flux in the atoms in the cell. After the valve has been opened, Rb atoms from the appendix diffuse through the tube, enter the cell and, after some bounces in the cell, are absorbed by the coating surface. Then, these atoms diffuse through the coating towards to the surface of the glass substrate, where they become irreversibly trapped.

We assume that the loss of atoms due to the chemical reaction of the atoms in the cell is negligible. Hence, in a steady-state the density of the atoms in the volume of the cell reaches an equilibrium value $n$ when the flux of atoms $I_{flux}$ from the appendix into the cell equals the rate of loss of the atoms on the glass substrate $F_{glass}$:

$$I_{flux} = F_{glass} \qquad (1)$$

$$S_{tube} D_{tube} \, grad \, n_{tube} = S_{coat} D_{coat} \, grad \, n_{coat} \qquad (2)$$

where $S_{tube} = \pi r^2$ is a cross-section of the cylindrical tube, $D_{tube}$ – diffusion coefficient for Rb atoms in the tube, $grad \, n_{tube} = (n_0 - n)/L$ is the atomic gradient in the tube. $D_{coat}$ - the diffusion coefficient of the atoms in the paraffin coating, $grad \, n_{coat} = n/l$ is the atomic gradient in the coating, $S_{coat} = 4 \pi R^2$ - the internal surface area of the cell. Taking this into account we can write:

$$S_{coat} \frac{n}{l} D_{coat} = S_{tube} \frac{n_0 - n}{L} D_{tube} \qquad (3)$$

In Knudsen's regime, the tube diffusion coefficient $D_{tube} = \frac{2r\bar{\upsilon}}{3}$ and the equilibrium vapor density $n$ can be written as follows:

$$n = \frac{n_0}{1 + \frac{6 D_{coat} \cdot R^2 \cdot L}{\bar{\upsilon} \cdot r^3 \cdot l}} \qquad (4)$$

where $\bar{\upsilon} = \sqrt{\frac{8kT}{\pi m}}$ is the mean thermal velocity at temperature $T$ and m is the mass of Rb atom.

Eq. 4 shows that, in the case of a thin coating when $6 D_{coat} \cdot R^2 \cdot L \ll \bar{\upsilon} \cdot r^2 \cdot l$, the density $n$ is proportional to the thickness $l$ of the cell. Conversely, with a thick coating when $6 D_{coat} \cdot R^2 \cdot L \ll \bar{\upsilon} \cdot r^2 \cdot l$, it saturates and at $l \to \infty$ it approaches to the maximum $n_0$. It can also be seen that, in order to optimize $n$, a cylindrical tube with a large diameter and a short length have to be made.

## 3. Optical pumping of atoms in a spherical bulb

Let us consider an optical pumping process in the bulb with an equilibrium vapor density $n$ of atoms. After the laser pump beam in the bulb volume has been switched on, the density of the optically polarized atoms starts to increase. In general, the polarization of the atoms offers three possibilities for them to be lost, namely: the escape of polarized atoms from the bulb through the cylindrical tube toward the appendix with Rb $\phi_{esc}$, the loss of the polarized atoms because of their depolarized collisions with the coating surface $\phi_{depolar}$ and the loss of the polarized atoms that were irreversibly adsorbed by the surface $\phi_{adsorb}$. The density of the pumped atoms in the cell reaches an equilibrium value $n_{polar}$ when the sum of all the loss rates of these atoms becomes equal to the production rate of the polarized atoms in the cell $P$ by the pump laser light:

$$P = p \cdot n \cdot V = \phi_{esc} + \phi_{depolar} + \phi_{adsorb} \tag{5}$$

where $p$ is a production rate of a single polarized atom that depends to the laser power, $V = 4\pi R^3 / 3$ is the bulb volume.

In a cell without a buffer gas, $\phi_{esc}$ can be calculated from the Knudsen's conductance of the cylindrical tube $K$ multiplied by the density gradient from the volume of the bulb to the appendix:

$$\phi_{esc} = K \cdot \frac{n_{polar}}{L} = \frac{2\pi r^3 \bar{\upsilon} n_{polar}}{3L} \tag{6}$$

Here we ignore the loss of the polarized atoms due to their depolarized collisions with the inner surface of the tube.

The loss rate $\phi_{depolar}$ due to depolarized collisions with the bulb coating can be expressed in the following form:

$$\phi_{depolar} = \frac{\pi R^2 \bar{\upsilon} n_{polar}}{\chi_{depolar}} \tag{7}$$

where $1/\chi_{depolar}$ is the probability that pumped atom to be depolarized by a single collision with the coating.

The loss rate $\phi_{adsorb}$ of pumped atoms due to their absorbing onto the bulb coating can be expressed as follows:

$$\phi_{adsorb} = \frac{\pi R^2 \bar{\upsilon} n_{polar}}{\chi_{adsorb}} \tag{8}$$

where $1/\chi_{adsorb}$ is the probability of polarized atom to be absorbed onto coating surface by a single collision.

The equilibrium condition (Eq. 3) becomes

$$P = \frac{2\pi r^3 \bar{\upsilon} n_{polar}}{3L} + \frac{\pi R^2 \bar{\upsilon} n_{polar}}{\chi_{depolar}} + \frac{\pi R^2 \bar{\upsilon} n_{polar}}{\chi_{adsorb}} \tag{9}$$

and the total number of polarized atoms in the cell $N_{polar} = n_{polar} V$ at equilibrium can be written as follows:

$$N_{polar} = P \left[ \frac{1}{\tau_{esc}} + \frac{1}{\tau_{depolar}} + \frac{1}{\tau_{adsorb}} \right]^{-1} = P \cdot \tau \tag{10}$$

where $\tau$ is the life time of the pumped atoms in the cell. It can be written as follows:

$$\tau = \frac{\chi \cdot l_{mean.path}}{\bar{\upsilon}} \tag{11}$$

where $l_{mean.path} = 4R/3$ is the mean path of pumped atoms in a spherical bulb. The parameter $\chi$ is the average number of bounces it takes to lose pumped atoms in the cell. Both the life-time and number of bounces are commonly measured in experiments and they characterize the quality of resonant cell as a whole.

Using Eq. 9 we express the life time as follows:

$$\tau = \left( \left( \frac{2L}{\bar{\upsilon}} \left( \frac{R}{r} \right)^3 \right)^{-1} + \left( \frac{\chi_{depolariz} l_{mean.path}}{\bar{\upsilon}} \right)^{-1} + \left( \frac{\chi_{adsorb} l_{mean.path}}{\bar{\upsilon}} \right)^{-1} \right)^{-1} \quad (12)$$

The first term in Eq. 12 is the inverse escape time $\tau_{esc}$ that represents the average time it takes to lose pumped atoms through the tube onto the piece of Rb in the appendix. It describes the "reservoir effect" discussed in detail in many publications (see, for example, [27]).

The second term is the inverse storage time $\tau_{depolar}$ of the atoms inside an isolated bulb before being lost due to depolarized collisions on the bulb walls: the parameter $\chi_{depolar}$ is interpreted as the average number of bounces it takes to depolarize atoms in a spherical bulb. The average number of bounces characterizes the quality of the coating itself to preserve long coherence times.

The third term is the inverse time $\tau_{adsorb}$ that represents the average time it takes to lose pumped atoms by adsorption onto the surface of the coating, The parameter $\chi_{adsorb}$ is interpreted as the average number of bounces it takes to adsorb atoms onto the coating,

Finally, we express both the total number of polarized atoms in the bulb in the steady-state regime as follows:

$$N_{polar} = p \cdot V \cdot n \cdot \tau \quad (13)$$

Eq. 13 shows that, in order to maximize the number of polarized atoms, the product of $n \cdot \tau$ should be kept as large as possible. This can be done, for instance, by maximization of $\tau$. To do so, a long and narrow tube have to be made in order minimize the reservoir effect. We note that this requirement for the maximization of $\tau$ is in a stark contradiction to the requirement, of the maximization of the equilibrium density $n$ of the atoms in bulb (see Eq. 4 ), that, conversely, requires a short and wide tube.

## 4. Experimental setup

The main part of the setup (see Fig. 1) is a cell made of a Pyrex glass bulb 1 (radius R) and a cylindrical tube 2 (internal radius $r$ and length $L$), connecting through a small valve 3 to an appendix 4 containing a piece of natural isotopic mixture of Rb. The bulb is connected through valve 5 to pump tube 6 with a vacuum gauge 7, then to valve 8 and finally to the vacuum turbo and ionic pumps that provide a vacuum up to $10^{-8}$ mbar. The densities ($n$ and $n_0$) of the Rb

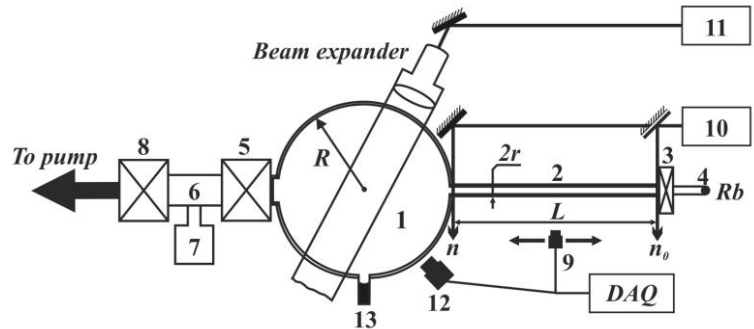

Fig. 1. 1- bulb; 2 - cylindrical tube; 4 - appendix with Rb; 3,5,8 - valves; 6 - pump tube, 7 - vacuum gauge; 8 - moveable photodiode; 10,11 - diode lasers, 12 - photodiode, 13 - container

vapor in the cell were measured by optical absorption through the two points of the tube. is the process was measured by a movable photodiode 9 with amplifier. The amplifier, in turn, was connected to a data acquisition system (DAQ). The light was delivered from a free-running diode laser 10. To eliminate the influence of optical pumping on the measurement of the density, the output power was attenuated down to several microwatts. The absolute vapor density $n_0$ at the source was estimated from the temperature of the of Rb metal drop [26]. In certain experiments on optical pumping, the fluorescence signal in the bulb (that is excited by a separate free-running pump laser 11) has been recorded by photo-detector 12 and a DAQ system. Laser 11 has a 5 mW power with an expanded beam diameter of 10 mm. Both lasers have linear polarized beams.

The dimensions of all parts of the cell have been carefully measured - the bulb has an internal diameter $2R = 12$ cm; the internal diameter the cylindrical tube is $2r = 0,8$ cm and its length $L = 10$ cm. The inner surface of the bulb, the cylindrical tube and the parts of the valves 3, 5 directed towards the bulb were covered by the paraffin film. We use paraffin which consists of short chains of wide-range molecular weights. It has a low melting temperature ranging from 65 - 75 °C. The coating was made by distilling of paraffin from a small container 13 (volume 90 mm$^3$) in the bottom of the bulb. This was done in a cell that had been thoroughly cleaned of all traces of water and oxygen by continuous pumping for several days and by a RF discharge of Ne. We consider the inner surface of the cell to have been cleaned when the discharge luminescence starts to assume a bright neon color and this color does not change remarkably after several hours of discharging. To make a paraffin film of desirable thickness, we heated container 13 for a calibrated period of time. The thickness was carefully determined by a measured volume of paraffin, distilled from the container, divided over the area of the whole coated surface. It was also possible to remove the coating by a separate heating of the bulb that then allows the paraffin to leak and distill back and distill freely into the cool container. Note, that the method described makes it possible to produce clean paraffin films of different thicknesses and to do so many times without altering the vacuum and other conditions in the cell.

## 5. Density of atoms in the paraffin-coated vapor cells

In this section, we present the result of a study of the density of Rb atoms in the cell. First of all, we pumped the cell continuously to obtain a residual-gas pressure of $10^{-8}$ mbar. Maintaining a continuous pumping of the cell, we heated the container and produced a film with a chosen thickness. After one day of pumping of the freshly coated cell, we closed valve 5, opened valve 3 of the vapor source and filled the bulb with Rb vapor.

To measure the steady-state density of Rb atoms $n$ in the bulb and the density at the vapor source $n_0$, we tuned the frequencies of laser 10 to be resonant with the optical transition of the Rubidium, until we obtained maximum light absorption in a separate not coated reference cell. We usually took one measurement for each thickness of the film, while five measurements were carried out for the thickness of 2 μm. Figure 2 shows the typical dependence of the density on the paraffin film thickness.

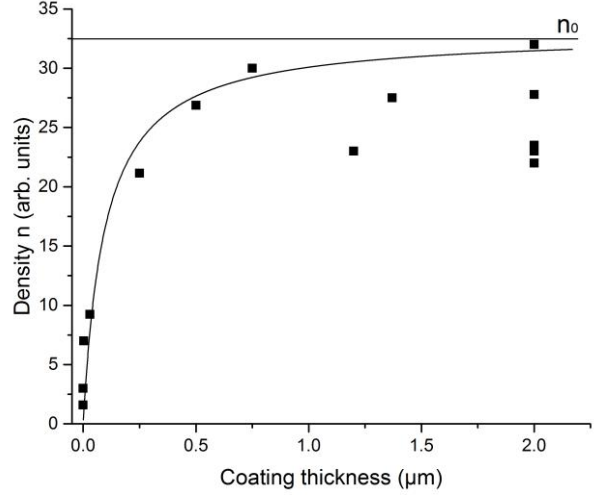

Fig. 2: Dependence of the equilibrium density $n$ on the paraffin film thickness. The curve is a best fit using Eq. 4.

It can be seen that the density in the bulb without a coating is close to zero, after which the density increases linearly as the thickness increases and it tends to saturate at a thickness of about of 2 μm. The form of the graph agrees with model (Eq. 4).

The experimental graph presented can be clearly divided into two parts: the first from zero thickness to a thickness of about 0.75 μm; the second from about 1 μm to 2 μm. In the first part of the graph, the equilibrium atomic density is a more or less smooth function of the thickness of the coating. In the second part, as can be seen, the density is a chaotic function of the thickness. A visual inspection of the coating revealed that a thin coating is very uniform, smooth and hardly visible. In contrast, a thick coating is fairly rough, milky colored and it is partially covered by small paraffin crystals that are visible even to the naked eye. It might be postulated that this roughness and the crystals provoke the generation of tiny cracks in the coating, through which atoms can penetrate close to the glass substrate surface and be lost onto it. Moreover, it was found that the surface imperfection of a thick coating leads to the formation of localized Rb metallic filaments (the so-called "whiskers" as is discussed in [28]). The effect of filament generation can be greatly accelerated by an increasing of atomic flux from by separate heating the Rb vapor source or by the cooling of the bulb. The whiskers can also be removed by gentle heating of the bulb up to the temperature of 40 $^0$C.

Due to this extra loss, the equilibrium density in the bulb with a thick coating is smaller than $n_0$ and it depends randomly to the film thickness because of the uncontrollable quality of the film. We attempted to make an improvement to the film quality using a slow distillation of paraffin but had not satisfactory results. This is because the process of film-growing by paraffin evaporation is naturally highly unstable: small local increases in the thickness of the film always have a tendency to increase even more during the process evaporation.

We use the plot presented in Fig.2 to determine the diffusion coefficient of the Rb atom in the paraffin film. This is achieved by fitting Eq. 5 to the experimental points in the first part of the plot only. The average value of the diffusion coefficient of Rubidium atoms $D_{coat}$ was found to be equal to $4,7 \cdot 10^{-7}$ cm$^2$/s. The statistical uncertainty is ± 5%, mainly attributable to a large amount of uncertainty in the measurements of the thicknesses of the coating. We found that this surprisingly large diffusion coefficient measured in our experiment is equal within an error bar to the paraffin (n-Octadecane) self - diffusion coefficient $4,6 \cdot 10^{-7}$ cm$^2$/s, that was measured in [29].

## 6. Relaxation of polarized atoms in paraffin-coated vapor-cells

To characterize the quality of the paraffin used, we performed measurements of the time relaxation polarization of the ground state of the Rb atoms $\tau_{depolar}$ and the average number of bounces $\chi_{depolar}$. Before all experiments, care was taken to assure that fluorescence was measured in an optically thin regime, hence the fluorescence intensity was always proportional to the vapor density. First we measured $\tau$. This is done as follows. Maintaining a continuous pumping of the cell, we heated container 9 and produce a coating with a thickness of one micrometer. We cooled the container down to room temperature, closed valve 5, opened valve 3 of the vapor source and filled the bulb with Rb vapor. The laser frequency of the pump laser 11 was tuned to be resonant to $5S_{1/2}$, F = 3 to $5P_{3/2}$, F′ = 4 of $^{85}$Rb transition, until a maximum of fluorescent intensity in the separate, not coated cell was obtained. Then the pump beam in the bulb was abruptly opened by a shutter. The atoms on the Rb atom ground state $5S_{1/2}$, F = 3 (state 1) were excited by the radiation; and the radiation populated $5S_{1/2}$, F = 2 ground state (state 2) through the intermediate atomic upper states $5P_{3/2}$, F' = 2, F' = 3, F' = 4. As a result of the optical pumping process, the vapor in the bulb became more transparent to the radiation, hence the vapor fluorescence dropped to a minimum. Then the beam was blocked. After a variable period of time $T$, the beam was opened again and the whole fluorescence signal was recorded.

Figure 3 shows a typical fluorescence signal. One can see that after the beam at zero time has been turned on, both stray light and fluorescence signals increased abruptly, then, due to optical depopulation of state 1, the fluorescence exhibited an extremely fast decay. We pointed out that, just after the beam was opened, no fluorescence signal was detected, except for the stray light signal $S_0$. At time 1,1 second, the laser light was turned off to allow the population from the state partially to return to state 1; and after a delay $T$, the light was switched on again. It could be observed that, at the time 1,6 seconds, both stray light and fluorescence signals re-appeared, but the amplitude of the second peak was smaller than the amplitude of the first. The difference between the two amplitudes corresponded to a non-complete relaxation polarization between both ground states caused by $T$ not being long enough.

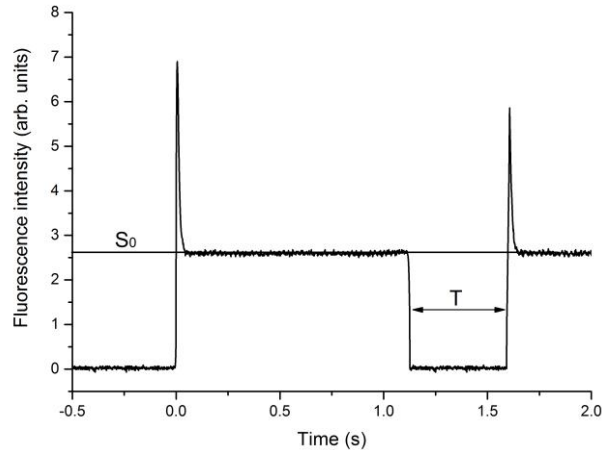

Fig. 3: Fluorescence signal

We extracted life-time $\tau$ from the dependence of amplitude of the second peak to the delay time $T$. The life-time was found to be equal to 0,42 s. For this value of life-time, bulb radius R = 6cm, $l_{mean.path} = 4R/3 = 8$, Rb thermal velocity $\bar{\upsilon} = 2,7 \cdot 10^4$ cm/s, Eq. 11 yielded the number of bounces $\chi \sim 1,4 \cdot 10^3$. Both life-time $\tau$ and number of bounces $\chi$ measured in this experiment characterize the quality of our cell.

Using this technique, it is not possible to measure either the escape time or the adsorption time of the polarized atoms in the cell. In order to measure these quantities, we used the method of suddenly introducing atoms into the cell. This was done with the help of the particles photo-desorption effect [18, 22, 30]. In the experiment, we postulated that the atoms and polarized atoms escaped or were be being absorbed onto the coating with identical times.

We pumped the cell, closed valve 5 and cooled the appendix by liquid nitrogen, leaving valve 3 open. With valve 3 open, photo-desorbed atoms were able to escape from the bulb into the cold appendix and be condensed there. Because we used a non-stabilized laser 11, we caused the laser frequency to sweep periodically across the Rb optical transitions by the triangle modulation of the laser pump current. This method of measuring the density allowed us to avoid any influence exerted by the frequency instability of the laser radiation on the measurements. The density was extracted from averaged amplitudes of the fluorescence of all Rb atom spectral lines.

At time t = 0, we illuminated the bulb by a photographic flash lamp. The flash lamp created a burst of rubidium atoms in the bulb volume by a pulse photo-desorption of the atoms from the coating. Then, we recorded the fluorescence decay that was due to the decreasing of the density of Rb atoms. Fig. 4 shows the dependence of the fluorescence signal of photo-desorbed atoms in the bulb as a function of time. One can see that the intensity of the fluorescence of the Rb atoms before the flash (sharp peak at t = 0) is close to the modulated intensity of the laser stray light.

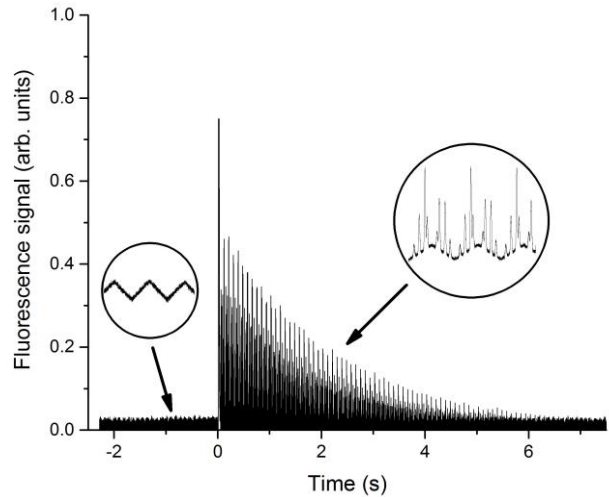

Fig. 4: Fluorescence signal as a function of time detected in the bulb with closed valve 5, opened valve 3 and cold appendix. The left insert shows modulated stray light of the laser, while the right one shows the same light together with series of Rb spectral lines. The spike at t=0 s is due to the photographic flash light hitting the photodetector

During the flash, the fluorescence intensity increases sharply and then, after approaching a maximum, the intensity slowly declines. Using the decay of the fluorescence intensity we measured the time of the density decay $\tau_{esc.adsorb}$ that was caused by both the escaping and adsorbing processes of the atoms. It was found that this time is equal to 2 s.

Then we measured the adsorption time $\tau_{adsorb}$. To do so we closed both valves 3 and 5 and completely illuminated an isolated bulb with the flash lamp. We recorded the florescence decay caused by the adsorbing of the atoms onto the coating and found $\tau_{adsorb}$ to be equal to 5s.

For bulb radius $R = 6$ cm, mean path $l_{mean.path} = 4R/3 = 8$ cm and $\tau_{adsorb} = 5$ s, the third term in Eq. 11 yields $\chi_{adsorb} \sim 1,7 \cdot 10^4$ s. By using the following equation

$$\tau_{esc} = \left(\tau^{-1}_{esc,adsorb} - \tau^{-1}_{adsorb}\right)^{-1} \tag{13}$$

and the measured parameters $\tau_{esc.adsorb}$ and $\tau_{adsorb}$ we evaluated that $\tau_{esc}$ was found to be equal to 3,3 s. This time is in agreement with the prediction given by the model. In fact, for bulb radius R = 6cm, tube dimensions $r = 0,4$ cm and $L = 10$ cm, Rb thermal velocity $\bar{\upsilon} = 2,7 \cdot 10^4$ cm/s, the first term in Eq. 12 yields $\tau_{esc} = 3,6$ s.

Taking into account the results of the measurements of life-time, escape time and adsorbing time, we estimated the relaxation polarization time $\tau_{depolar}$ and the average number of bounces $\chi_{depolar}$ for the paraffin used. For the values: $\tau = 0,42$s, $\tau_{esc} = 3,3$s and $\tau_{adsorb} = 5$s Eq. 10 yields $\tau_{depolar} = 0.53$s. Finally, for the measured $\tau_{depolar} = 0.53$s, the second term in Eq. 11 yields $\chi_{depolar} \sim 2 \cdot 10^3$.

The results of measurements are summarized in Table 1

Table 1

| $\tau$ | $\chi$ | $\tau_{depolar}$ | $\chi_{depolar}$ | $\tau_{esc}$ | $\tau_{esc}$ (model) | $\tau_{adsorb}$ | $\chi$ |
|---|---|---|---|---|---|---|---|
| 0,42 s | $1,4 \cdot 10^3$ | 0.53 s | $2 \cdot 10^3$ | 3,3 s | 3,6 s | 5s | $1,7 *10^4$ s |

It can be seen from the Table 1 that it takes $1,7 \cdot 10^4$ collisions for the Rb atoms to be irreversibly absorbed into the paraffin coating. This result is consistent with the values published in the literature for various alkalis, preponderantly Na on paraffin [31] and Rb on polydimethilsiloxane (PDMS) coating [32]. The measured number shows that the probability for Rb atom to relax on the paraffin surface in one collision ($1/\chi_{depolar} = 5 \cdot 10^{-4}$) is about one order of magnitude large than the probability for the atom to be irreversibly adsorbed into the paraffin coating ($1/\chi = 6 \cdot 10^{-5}$).

As mentioned above, we ignored the loss of the polarized atoms due to their depolarized collisions with the inner surface of the tube. By using the average numbers measured, it is possible to estimate the distance $X$ that polarized Rb atoms need to penetrate inside the tube and to relax completely on the paraffin surface. The distance can be written as [33]:

$$X = d\sqrt{2\chi_{depolar}/3} \tag{14}$$

For the average number $2 \cdot 10^3$ and tube internal diameter $d = 0,8$ cm, the equation (14) yields $X \sim 45$ cm that is longer than the tube length (10 cm). One can conclude that, in our case, the polarized atoms were lost in the tube mainly due to their collisions with the Rb metal surface in the appendix.

From the data presented, one can see that the serious limitation of the equilibrium numbers of atoms and the life-time of the polarized atoms in the cell for a given coating is the escape of the atoms towards to the vapor source. As can be seen from Eq. 12, the escape time is proportional to the length $L$ of the atomic source tube, and to the cube of the cell radius $R$ divided by the cube of the tube radius $r$. This means that both escape rate of the polarized atoms and reservoir effect that are relatively insignificant in big cells become much more important in small cells. For example, for a small spherical bulb with radius 0,1 cm, tube radius 0,01 cm and length of 0,2 cm the escape time is equal to 15 ms. For the present best coating [15] with $\chi_{depolar} = 10^6$ one can estimate $\tau_{depolar}$ to be equal 5 s, and the life-time $\tau$ for the small cell is equal to 14 ms.

As mentioned above, the suppression of the escape polarized atoms requires the use of a long and narrow tube. This is in a contradiction to the requirement that the maximization of the atomic equilibrium density in the bulb requires a short and wide tube.

**Acknowledgements**

We would like to thank Prof. A. Shalagin, Prof. N. Surovtsev for stimulating discussions. Special thanks for A. Makarov for the help in experiments and R. Robson (McKillop) for careful reading of the manuscript. The work is supported by Russian Foundation for Basic Research (RFBR), grant number 15-02-02333 A